\def\year{2021}\relax
\documentclass[letterpaper]{article} 
\usepackage{aaai21}  
\usepackage{times}  
\usepackage{helvet}  
\usepackage{courier}  
\usepackage{url}  
\usepackage{graphicx}  
\usepackage{float}
\usepackage{multirow}
\usepackage{diagbox}
\usepackage{amsmath}
\usepackage[linesnumbered,titlenumbered,ruled,vlined,commentsnumbered,noend]{algorithm2e}

\usepackage{times}
\usepackage{xcolor}
\usepackage{soul}
\usepackage[utf8]{inputenc}
\usepackage[small]{caption}
\usepackage{comment}
\usepackage{color}

\usepackage{epsfig}
\usepackage{graphicx}
\usepackage{amsmath}
\usepackage{amssymb}
\usepackage{natbib}

\usepackage{url}
\usepackage{multirow}
\usepackage{enumitem}
\usepackage{moresize}
\usepackage{epstopdf}
\usepackage{array}

\usepackage[switch]{lineno}  %
\usepackage{booktabs}
\usepackage{wrapfig}
\usepackage{subfig}
\usepackage[hang,flushmargin]{footmisc} 

\usepackage{xspace}

\usepackage[colorlinks,
linkcolor=red,
citecolor=blue
]{hyperref}

\usepackage{xspace}
\makeatletter
\DeclareRobustCommand\onedot{\futurelet\@let@token\@onedot}
\def\@onedot{\ifx\@let@token.\else.\null\fi\xspace}
\def\eg{\emph{e.g}\onedot} 
\def\ie{\emph{i.e}\onedot} 
 
\def\etc{\emph{etc}\onedot} 
 
\def\etal{\emph{et al}\onedot}
\makeatother

\definecolor{darkorange}{rgb}{1.0, 0.55, 0.0}
\definecolor{lincolngreen}{rgb}{0.11, 0.35, 0.02}
\definecolor{cornflowerblue}{rgb}{0.39, 0.58, 0.93}
\definecolor{cobalt}{rgb}{0.0, 0.28, 0.67}

\usepackage{threeparttable}

\title{Generating Adversarial Examples with Controllable Non-Transferability}

\author{AAAI 2021 Anonymous Submission \#4389}

\author {
    Renzhi Wang\textsuperscript{\rm 1}, 
    Tianwei Zhang\textsuperscript{\rm 2},
    Xiaofei Xie\textsuperscript{\rm 2}, 
    Lei Ma\textsuperscript{\rm 3}, 
    Cong Tian\textsuperscript{\rm 4}, 
    Felix Juefei-Xu\textsuperscript{\rm 5}, 
    Yang Liu\textsuperscript{\rm 2} \\
}

\affiliations {
    \textsuperscript{\rm 1} Intel Corporation, China \\
    \textsuperscript{\rm 2} Nanyang Technological University, Singapore \\
    \textsuperscript{\rm 3} Kyushu University, Japan \\
    \textsuperscript{\rm 4} Xidian University, China \\
    \textsuperscript{\rm 5} Alibaba Group, USA \\
}

\begin{document}
\maketitle

\begin{abstract}
Adversarial attacks against deep neural networks have been widely studied. One significant feature that makes such attacks particularly powerful is \emph{transferability}, where the adversarial examples generated from one model can be effective against other similar models as well. A large number of works have been done to increase the transferability. However, how to decrease the transferability and craft malicious samples \emph{only} for specific target models are not explored yet. 
In this paper, we design novel attack methodologies to generate adversarial examples with controllable non-transferability. With these methods, an adversary can efficiently produce precise adversarial examples to attack a set of target models he desires, while keeping benign to other models. The first method is \emph{Reversed Loss Function Ensemble}, where the adversary can craft qualified examples from the gradients of a reversed loss function. This approach is effective for the white-box and gray-box settings. The second method is \emph{Transferability Classification}: the adversary trains a transferability-aware classifier from the perturbations of adversarial examples. This classifier further provides the guidance for the generation of non-transferable adversarial examples. This approach can be applied to the black-box scenario. Evaluation results demonstrate the effectiveness and efficiency of our proposed methods. This work opens up a new route for generating adversarial examples with new features and applications.
\end{abstract}

\section{Introduction}\label{sec:intro}

The past decade has witnessed the revolutionary development of Deep Learning (DL) technology. A variety of DL algorithms and Deep Neural Network (DNN) models are designed to enable many artificial intelligent tasks, especially in the safety- and security-critical domains, \eg, autonomous driving, security authentication, intrusion detection, \etc. As a result, DL models are expected to meet high requirements of robustness and security. 
Unfortunately, DNNs are well-known to be vulnerable to adversarial examples (AEs): with imperceptible modifications to the input sample, the model will be fooled to give wrong prediction results. Since its discovery in 2013 \cite{szegedy2013intriguing}, a large amount of efforts have been spent to launch and heat the arms race between the adversarial attacks and defenses.


Adversarial examples enjoy one important feature: \emph{transferability} \cite{szegedy2013intriguing}. AEs generated from one model has certain probability to fool other models as well.
Due to this transferability, it becomes feasible for an adversary to attack black-box models, as he can generate AEs from an alternative white-box model and then transfer it to the target model. As a result, extensive works have been focusing on increasing AE's transferability \cite{Dong2018Boosting, dong2019evading, xie2019improving}, explaining the mechanism and effects of transferability \cite{tramer2017ensemble,athalye2018Obfuscated,wu2018understanding,zhang2020towards}, and trying to attack in real world\cite{guo2020abba}.

Different from those studies, we aim to explore how to decrease the transferability and generate AEs only for certain DL models desired by the adversary. Generating AEs with such controllable non-transferability can enable the adversary to precisely attack his targets without affecting other irrelevant or his own applications. For instance, Autonomous Vehicles (AVs) adopt state-of-the-art DNN models for objection recognition and motion planning. A malicious AV vendor may want to attack its competitor's products. He can generate AEs only for the target model, and physically install them in the real-world scenarios. Then the vehicles from its rival company will malfunction when meeting the AEs, while its own vehicles can behave correctly.

\begin{figure*}
    \centering
    \includegraphics[width=0.85\linewidth]{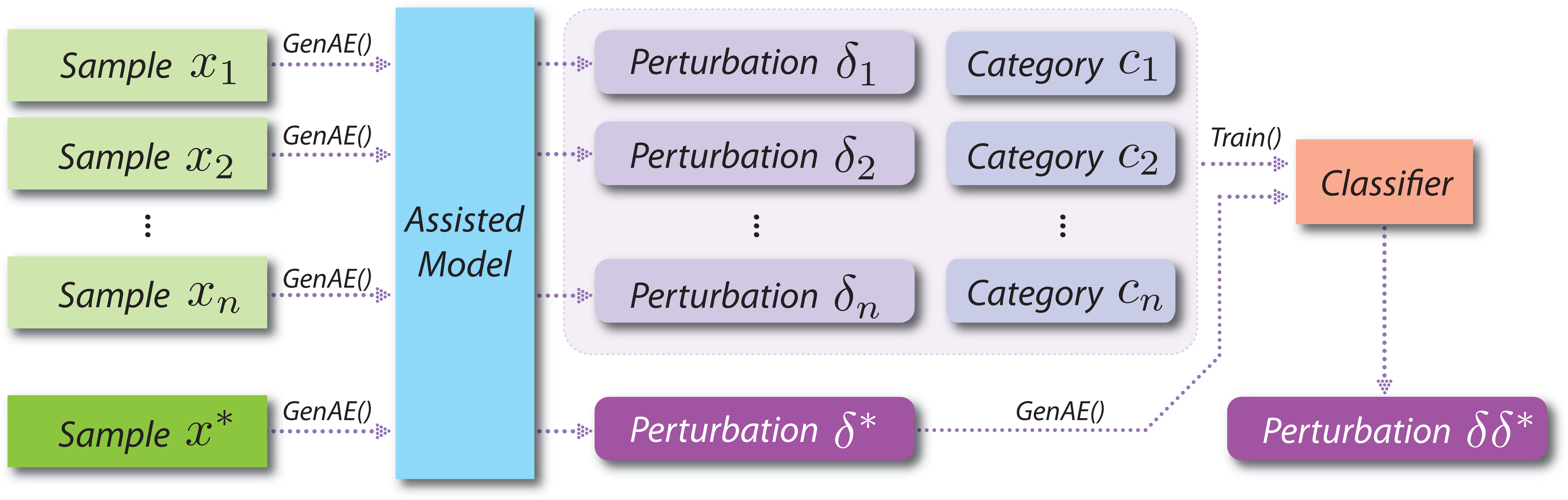}
    \caption{Overview of our Transferability Classification methodology. We generate adversarial examples by an assisted model first. And then train transferability-aware classifiers. Finally we perform re-attack on these classifiers, to generate non-transferable AE in black-box attack.}
    \label{fig:reattack}
\end{figure*}

There are two challenges towards this goal. First, transferability is a nature feature for AEs. It is hard to efficiently craft AEs to minimize their damages on specific protected models, while still maintaining high success rate on specific target models. Second, we need to consider different attack settings. When the adversary has little knowledge about the details and parameters of target or protected models, it is also difficult for him to generate accurate AEs to achieve his goal. To the best of our knowledge, there is currently only one study about this non-transferability feature \cite{nakkiran2019a}. However, this work only considers the white-box setting. The generated AEs are not controllable: the attack aims to decrease the transferability over all models instead of some specific protected ones. Beside, it is unknown whether this approach is applicable to models with different structures.

\begin{figure}[!ht]
  \centering
  \subfloat[Origin Image.\newline Resnet50:airplane \newline VGG16:airplane]{\includegraphics[width=0.3\columnwidth]{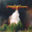}}
  \quad
  \subfloat[Origin Resnet50 Explanation Region]{\includegraphics[width=0.3\columnwidth]{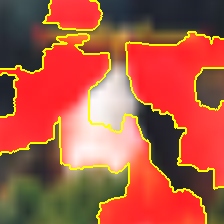}}
  \quad
  \subfloat[Origin VGG16 Explanation Region]{\includegraphics[width=0.3\columnwidth]{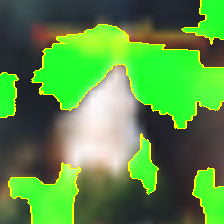}}
  \\
  \subfloat[Normal AE \newline Resnet50:ship \newline VGG16:ship]{\includegraphics[width=0.3\columnwidth]{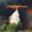}}
  \quad
  \subfloat[Normal \newline Resnet50  (attack)\newline Explanation Region]{\includegraphics[width=0.3\columnwidth]{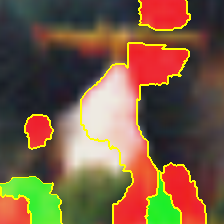}}
  \quad
  \subfloat[Normal \newline VGG16 (protect) \newline Explanation Region]{\includegraphics[width=0.3\columnwidth]{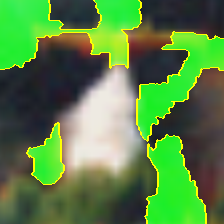}}
  \\
  \subfloat[AE with \newline Non-transferability \newline  Resnet50:ship \newline VGG16:airplane]{\includegraphics[width=0.3\columnwidth]{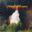}}
  \quad
  \subfloat[Non-transfer \newline  Resnet50 (attack)\newline Explanation Region]{\includegraphics[width=0.3\columnwidth]{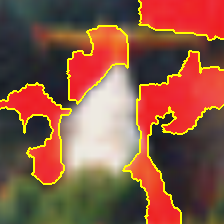}}
  \quad
  \subfloat[Non-transfer \newline VGG16 (protect) \newline Explanation Region]{\includegraphics[width=0.3\columnwidth]{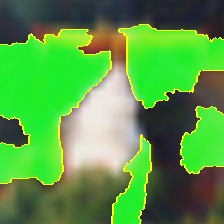}}
  \\
  
  \caption{The example of the comparison between normal adversarial examples and adversarial examples with non-transferability. Compare to normal AE attack, our method can protect specific model from attack, the explanation area is very close to the origin image, while the other AE images' explanation areas are far more different with origin image. The explanation areas are generated by Lime \cite{ribeiro2016why}}
  \label{fig:samples}
\end{figure}

In this work, we aim to generate adversarial examples with controllable non-transferability like Fig.~\ref{fig:samples}. For this goal, we propose two attack methodologies the first one is \emph{Reversed Loss Function Ensemble}, for the white-box or gray-box attack settings. We design a new loss function which considers the impacts of both the target models, and the protected models (in a reversed fashion). Optimization of this loss function can give very precise AEs with success rates of higher than $95\%$ on the target models, and lower than $1\%$ on the protected models. The second solution is \emph{Transferability Classification}, for the black-box attack setting. We find that the the adversarial perturbations with different transferability results have distinct characteristics. Based on this observation, we build a transferability-aware classifier, and generate AEs efficiently from it to meet our demands.

We have evaluated our methods on two widely-used datasets (\ie, CIFAR-10 and ImageNet) and three popular DNNs (\ie, ResNet50, VGG16 and DenseNet121). The evaluation results demonstrate that 1) under the white-box setting, our method could reduce the non-transferability to up to $\textbf{0.04\%}$, 2) under the gray-box setting, the non-transferability can be reduced to up tp $\textbf{5.26\%}$ and 3) under the black-box setting, the adversary can reduce the non-transferability across different models as low as $\textbf{12\%}$, even he has very limited knowledge about those models.

\section{Related Work}\label{sec:related}

\subsection{Generating AEs with High Transferability}
The transferability feature of AEs was first identified with the L-BFGS method in \cite{szegedy2013intriguing}. Since then, a number of works have been done to increase the transferability of the gradient-based methods for AE generation. MI-FGSM \cite{Dong2018Boosting} adopted momentum optimization in the process of gradient generation to enhance the transferability. This helps it win the 1st place in both targeted and untargeted adversarial attack competitions of NeurIPS2017. DI$^2$-FGSM \cite{xie2019improving} applied random transformations to the input at each iteration of AE generation to achieve high success rate on black-box models. TI-FGSM \cite{dong2019evading} enhanced the transferability of AEs towards the models with adversarial defenses. It identified the region which is less sensitive to white-box models but more sensitive to the target model, and adopted a pretrained matrix kernel to handle the gradients. These three methods are all based on FGSM, and can be combined to further improve the transferability. TAP \cite{zhou2018transferable} maximized the distance between clean samples and the corresponding AEs in the intermediate feature map, and adopted smooth regularization to improve the success rates on both white-box and black-box models. 

\subsection{Explanation and Understanding of Transferability}
There are different points of view to understand the mechanism of transferability. The first one is from the model decision boundary. Liu \etal \cite{liu2016delving} proposed that transferability is caused by the similarity of decision boundaries of different models, even they have different network structures. An AE that can cross the boundary of one model has certain probability to cross the boundaries of other models as well. Tramer \etal \cite{tramer2017space} extended this hypothesis with the concept of adversarial subspace. Models with a higher-dimensional adversarial subspace have higher probabilities to coincide, thus the transferability is higher. The second perspective is from the data features. Ilyas \etal \cite{ilyas2019adversarial} proposed the concept of non-robust features that can be leveraged for AE generation. Different models for the same task can utilize similar non-robust features. As a result, AEs constructed from those features can be transferred across each other. The third direction is to identify the factors that can affect the transferability. Wu \etal \cite{wu2018understanding} discovered that models with similar architecture, lower capability and high test accuracy have higher transferability. Papernot \etal \cite{papernot2016transferability} pointed out that transferability is not only limit to neural networks. They exist in statistic machine learning models (\eg, logistic regression, SVM, KNN, decision tree) as well.

\section{Method 1: Reversed Loss Function Ensemble}
\label{sec:white-gray-box}
\subsection{Attacking White-box Models}
\label{sec:whitebox}
We first consider a white-box setting, where the adversary has full knowledge of the models in $\mathbb{M}_{\mathrm{attack}}$ and $\mathbb{M}_{\mathrm{protect}}$. 
This scenario occurs when the model owners adopt the DL models published online for free use.
The adversary can calculate the gradients of the model parameters to generate precise non-transferable adversarial examples. 

Recall that the generation of adversarial examples for one target model $f$ can be formulated as an optimization problem in Equation \ref{eq:optimize}(a). 
To solve this problem, one typical approach is to maximize the loss function in Equation \ref{eq:optimize}(b), where $\langle f(x) \rangle_j$ denotes the confidence score of the $j$-th output, and $y$ denotes the correct label of $x$.   
\begin{subequations}
\label{eq:optimize}
  \begin{align}
& \mathrm{minimize~}  \|\delta\|, \mathrm{~~subject~to~~} \: f(x+\delta)\neq f(x) \\
& \mathrm{L}(f, x, y) = -\langle f(x) \rangle_{y}+\log{ \left(\sum_{j\neq y}{\exp{\langle f(x) \rangle_j}} \right)}
  \end{align}
\end{subequations}
We introduce Reversed Loss Function Ensemble (RLFE) to handle multiple models from the attacked model set $\mathbb{M}_{\mathrm{attack}}$ and protected model set $\mathbb{M}_{\mathrm{protect}}$. Specifically, we ensemble the loss function from each model into a combined function (Equation \ref{equation reversed ensemble}): the attacked models positively contribute to the total loss while the protected models have negative impacts on this function. By maximizing this loss function, the identified adversarial examples are guaranteed to mislead the attacked models while hardly affecting the outputs of the protected models. 
\begin{align}\label{equation reversed ensemble}
    \mathrm{L}^* = \sum_{f_i \in \mathbb{M}_{\mathrm{attack}}}{\lambda_i \cdot \mathrm{L}(f_i,x,y)} - \sum_{f_j \in \mathbb{M}_{\mathrm{protect}}}{\lambda_j \cdot \mathrm{L}(f_j,x,y)}
\end{align}

It is worth noting that RLFE is very accurate since the adversary has white-box accesses to the models' parameters and generates the precise perturbations on those models. So even the attacked and protected models have very minor differences, the adversary is still able to discover examples that can distinguish them. 
This is hard to achieve in gray-box or black-box scenarios, due to the limited adversarial capabilities. 

\subsection{Attacking Gray-box Models}\label{section Semi White-box Attack Method}

Next we consider the gray-box setting, where the adversary knows the training details and configurations of the models, but not the parameter values. This is the scenario when the model owner adopts the public DL algorithms or cloud services to train his own copy of models.

In this case, the adversary cannot directly employ RLFE to calculate the adversarial perturbations due to the lack of model parameters. Instead, he can first construct shadow models with the same algorithms and configurations to mimic the behaviors of the actual models. Then he can generate the adversarial examples using RLFE over the shadow models. These examples can be used to attack or protect the original models as well. Note that this solution can also be applied to the case where some of the models are white-box to the adversary while the rest are gray-box. The adversary only needs to construct the shadow models of the gray-box ones and use the parameters of white-box models directly when generating adversarial examples. 

It is interesting that we \emph{utilize the transferability between shadow and actual models to generate non-transferable AEs between attacked and protected models.} As a result, the quality of adversarial examples depends on two factors. The first factor is the similarity between the shadow models and actual models. A higher similarity can improve the transferability between the constructed shadow and actual models, making the results closer to the whit-box setting. Various model extraction techniques \cite{Tramer2016Stealing,wang2018stealing,yu2020cloudleak,jagielski2020high} have been proposed, which can be leveraged by the adversary to build the shadow models. The adversary can use the same or similar training set to train replicas of the actual model to achieve the same performance. He can also synthesize a dataset by querying the actual model with certain samples, and use the confidence scores as the output. This has been demonstrated more effective and efficient \cite{jagielski2020high}.

The second factor is the distinction between the attacked and protected models. If the attacked and protected models have very similar behaviors and features (\eg, originating from the same training algorithms and settings), it is hard to construct shadow models to differentiate them. This is one limitation in the gray-box setting, due to the restricted capability. The more distinct are the structures of the attacked models from the protected ones, the more effective our approach will be.

\section{Method 2: Transferability Classification}
\label{sec:blackbox}

Finally we consider the black-box scenario. Now the adversary does not have knowledge about the protected and attacked models, and cannot construct shadow models and then apply RLFE. We only assume the adversary can query those black-box models, \ie, sending arbitrary input and retrieving the corresponding output. One straightforward way is to keep generating adversarial examples and testing their transferability until a qualified one is discover. This brute force method is not efficient since for each sample, the adversary has to try a lot of sessions before obtaining a good one. Instead, we aim to build a generator, that can produce the desired adversarial perturbations corresponding to a given input easily and promptly. To achieve this goal, we propose a novel method, \emph{Transferability Classification}, which could precisely capture the transferability 
information between two models. 

It has been demonstrated that the transferability of AEs is caused by the common \textit{non-robust} features that are learned by different models~\cite{ilyas2019adversarial}. Inspired by this, we further find that the non-transferability could also be controlled by identifying some specific features learned by one model but not by others. The key insight of our solution is that \emph{the adversarial perturbations exhibit certain inherent characteristics that can represent their transferability across different models.} So we can build a transferability-aware classifier (TC) to identify the relationship and facilitate the generation of new perturbations based on our demands. Our approach consists of two steps: (1) training a TC; (2) generating non-transferable adversarial examples by attacking on this classifier. 
Fig.~\ref{fig:reattack} illustrates the methodology overview.

\subsection{Step 1: Training Transferability-aware Classifier}\label{section Noise-based Transfer Classifier}
Without loss of generality, we consider two models: one is attacked model ($f_{\mathrm{attack}}$) and another one is protected model ($f_{\mathrm{protect}}$). So there are four categories of adversarial examples: (I) succeed in attacking both models; (II) succeed in attacking $f_{\mathrm{attack}}$ but fail to attack $f_{\mathrm{protect}}$; (III) succeed in attacking $f_{\mathrm{protect}}$ but fail to attack $f_{\mathrm{attack}}$; (IV) fail to attack both models. The goal of this step is to train classifiers, such that given an adversarial perturbation, they are able to predict which type of attack results this perturbation belongs to without querying the models. 

\textbf{Dataset preparation.}
We need a dataset to train the TC. We prepare an assisted white-box model ($f_{\mathrm{assist}}$) which performs the same task as $f_{\mathrm{protect}}$ and $f_{\mathrm{attack}}$. Then we generate a large quantity of adversarial examples from $f_{\mathrm{assist}}$ using conventional methods (We use BIM \cite{kurakin2016adversarial} to generate AEs in this paper, but other methods can be used in the same way). For each example, we label them as one of the four categories defined above by querying $f_{\mathrm{protect}}$ and $f_{\mathrm{attack}}$. Then we are able to synthesize a four-class dataset with the added perturbation ($\delta$) as the feature and the attack result as the label. Note that this process can give unbalanced samples and make the classifier biased. We prune the dataset to guarantee each class has the same number of samples.

\textbf{Classifier training.}
We split this dataset as training set and validation set, and train classifiers over them. We adopt a DNN model, Resnet18 as the structure of our classifier. Other networks can be used as well. Detailed experimental configurations can be found in Section \ref{sec:setup}.  We consider two strategies: the first one is to train a unified classifier to predict the samples from all classes of the target models. The second one is to train a label-specific classifier for the samples from each class. Table \ref{table Noise Classifier accuracy} reports the training and testing accuracy of different combinations of white-box and target models. We can observe that each type of classifier can achieve a very satisfactory accuracy to differentiate four categories. The label-specific classifiers have better performance. So we will adopt this strategy in the follow evaluations. 

\begin{table}[!t]
    \centering
    \footnotesize

    \resizebox{1\linewidth}{!}{
    \begin{tabular}{l|l|l|c|cc}
        \toprule
        \textbf{White-box $f_{\mathrm{assist}}$} & \textbf{Target $f_{\mathrm{attack}}$} & \textbf{Target $f_{\mathrm{protect}}$} & \textbf{Classifier type} & \textbf{Train acc} & \textbf{Test acc}\\
        \midrule
        \multirow{2}{*}{Resnet50} & \multirow{2}{*}{VGG16BN} & \multirow{2}{*}{Densenet121} & Label-specific & $0.945$ & $0.906$\\ 
        ~ & ~ & ~ & Unified & $0.750$ & $0.748$ \\
        \midrule
        \multirow{2}{*}{VGG16BN} &  \multirow{2}{*}{Resnet50} & \multirow{2}{*}{Densenet121} & Label-specific & $0.968$ & $0.922$\\ 
        ~ & ~ & ~ & Unified & $0.793$ & $0.788$\\
        \midrule
        \multirow{2}{*}{Densenet121} & \multirow{2}{*}{Resnet50} & \multirow{2}{*}{VGG16BN} & Label-specific & $0.927$ & $0.922$\\ 
        ~ &~ & ~ & Unified & $0.793$ & $0.788$\\
        \bottomrule
    \end{tabular}
    }
        \caption{Training and testing accuracy of our Transferability-aware Classifier.}
    \label{table Noise Classifier accuracy}
\end{table}

\subsection{Step 2: Generating Adversarial Examples from Transferability-aware Classifier}

With the trained TC we can easily generate adversarial examples can only attack $f_{\mathrm{attack}}$ other than $f_{\mathrm{protect}}$. Specifically, given an input $x$, we use conventional methods to generate the corresponding perturbation $\delta$ which the TC can give correct prediction. If the perturbation can meet our demand (\ie, it belongs to category II), then we will select it as the final result. Otherwise, we treat this perturbation $\delta$ as the input sample of the TC, and use conventional methods to generate a second order of perturbation $\delta\delta$ that can shift the prediction result to category II. Then we are able to get the final result $x+\delta+\delta\delta$ to satisfy our goal. We can use this adversarial sample to query the two models for validation. We repeat the above process in case the validation indicates the generated AE does not work as expected. 

Note that once we obtain the TC, normally we need to run two generation sessions to identify the desired samples. The first session is to generate the perturbation $\delta$ from the assisted model $f_{\mathrm{assist}}$, which can converge in a few iteration. The second session is to generate the perturbation $\delta\delta$ from TC. This process needs dozens of iterations with a smaller $\epsilon$, but takes much shorter time due to the simple network structures of TC. In contrast, the brute force needs to run a large number of the first sessions to identify the satisfied AE. This makes our solution much more efficient. 

We use two models to describe our methodology. It can be extended to multiple models: if there are $n$ models in the attacked and protected model sets, we need to build TC with $2^n$ labels to distinguish different sorts of perturbations, and generate the perturbations in the category of attacking all attacked models and bypassing all protected models.

\begin{table}[!t]
\centering

\subfloat[][Cifar10]{
\label{table:whitebox-cifar10}
\resizebox{0.8\linewidth}{!}{
\begin{threeparttable}
\begin{tabular}{c|c|c|c}
  \toprule
\multicolumn{2}{c|}{Attacked Model} & \multicolumn{2}{c}{Protected Model} \\
Structure & Succ. Rate (\%) & Structure & Succ. Rate (\%) \\ 
    \midrule
 R50 & 99.61 $\rightarrow$ \textbf{99.55} & V16 & 74.78 $\rightarrow$ \textbf{1.24} \\
 \midrule
 R50 & 99.61 $\rightarrow$ \textbf{98.64} & D121 & 79.83 $\rightarrow$ \textbf{1.61} \\
 \midrule 
 \multirow{2}{*}{R50} & \multirow{2}{*}{99.61 $\rightarrow$ \textbf{98.36}} & V16 & 74.78 $\rightarrow$ \textbf{1.05} \\
 & & D121 & 79.83 $\rightarrow$ \textbf{1.59} \\
 \midrule
 V16 & 97.98 $\rightarrow$ \textbf{97.35} & R50  & 43.18 $\rightarrow$ \textbf{0.28}\\
 \midrule
 V16 & 97.98 $\rightarrow$ \textbf{95.44} & D121 & 46.13 $\rightarrow$ \textbf{0.31}\\
 \midrule
 \multirow{2}{*}{V16} & \multirow{2}{*}{97.98 $\rightarrow$ \textbf{93.99}} & R50 & 43.18 $\rightarrow$ \textbf{0.17}\\
 ~ & ~ & D121 & 46.13 $\rightarrow$ \textbf{0.14}\\
 \midrule
 D121 & 98.51 $\rightarrow$ \textbf{97.86} & R50 & 77.54 $\rightarrow$ \textbf{2.77}\\
 \midrule
 D121 & 98.51 $\rightarrow$ \textbf{98.53} & V16 & 79.60 $\rightarrow$ \textbf{2.50}\\
 \midrule
 \multirow{2}{*}{D121} & \multirow{2}{*}{98.51 $\rightarrow$ \textbf{98.88}} & R50 & 77.54 $\rightarrow$ \textbf{2.47}\\
 ~ & ~ & V16 & 79.60 $\rightarrow$ \textbf{1.85}\\
 \bottomrule

  \end{tabular}
  \end{threeparttable}
  }
  } \\
\subfloat[][ImageNet]{
\label{table:whitebox-ImageNet}
\resizebox{0.8\linewidth}{!}{
\begin{threeparttable}
\begin{tabular}{c|c|c|c}
  \toprule
\multicolumn{2}{c|}{Attacked Model} & \multicolumn{2}{c}{Protected Model} \\
Structure & Succ. Rate (\%) & Structure & Succ. Rate (\%) \\ 
 \midrule
 R50 & 99.87 $\rightarrow$ \textbf{99.63} & V16 & 50.28 $\rightarrow$ \textbf{0.58} \\
 \midrule
 R50 & 99.87 $\rightarrow$ \textbf{99.56} & D121 & 51.31 $\rightarrow$ \textbf{0.19} \\
 \midrule 
 \multirow{2}{*}{R50} & \multirow{2}{*}{99.87 $\rightarrow$ \textbf{99.26}} & V16 & 50.28 $\rightarrow$ \textbf{0.70} \\
 & & D121 & 51.31 $\rightarrow$ \textbf{0.20} \\
 \midrule
 V16 & 99.65 $\rightarrow$ \textbf{99.29} & R50  & 37.93 $\rightarrow$ \textbf{0.22}\\
 \midrule
 V16 & 99.65 $\rightarrow$ \textbf{99.22} & D121 & 39.20 $\rightarrow$ \textbf{0.11}\\
 \midrule
 \multirow{2}{*}{V16} & \multirow{2}{*}{99.65 $\rightarrow$ \textbf{98.89}} & R50 & 37.93 $\rightarrow$ \textbf{0.20}\\
 ~ & ~ & D121 & 39.20 $\rightarrow$ \textbf{0.13}\\
 \midrule
 D121 & 99.92 $\rightarrow$ \textbf{99.86} & R50 & 52.83 $\rightarrow$ \textbf{0.29}\\
 \midrule
 D121 & 99.92 $\rightarrow$ \textbf{99.83} & V16 & 51.42 $\rightarrow$ \textbf{0.05}\\
 \midrule
 \multirow{2}{*}{D121} & \multirow{2}{*}{99.92 $\rightarrow$ \textbf{99.75}} & R50 & 52.83 $\rightarrow$ \textbf{0.28}\\
 ~ & ~ & V16 & 51.42 $\rightarrow$ \textbf{0.04}\\
 \bottomrule

  \end{tabular}
  \end{threeparttable}
  }
  } 
  \caption{Attack success rates of different network structures under the white-box setting}
\label{table:white-box-attack}
\end{table}
\begin{table}[!t]
\centering

\subfloat[][Cifar10]{
\label{table:defense-cifar10}
\resizebox{0.75\linewidth}{!}{
    \begin{tabular}{c|c|c}
        \toprule
        \multirow{3}{*}{Structure} &Attacked Model & Protected Model \\
        ~ & (w/ adv. training) & (w/o adv. training) \\
        ~ & Succ. Rate (\%) & Succ. Rate (\%) \\ 
        \midrule
        R50 &99.74 $\rightarrow$ \textbf{98.31} & 18.25 $\rightarrow$ \textbf{0.05} \\
        \midrule
        V16 &98.93 $\rightarrow$ \textbf{98.77} & 28.00 $\rightarrow$ \textbf{0.41} \\
        \midrule 
        D121 &99.95 $\rightarrow$ \textbf{99.45} & 22.19 $\rightarrow$ \textbf{0.13}\\
        \bottomrule
  \end{tabular}}
} \\
\subfloat[][ImageNet]{
\label{table:defense-ImageNet}
\resizebox{0.75\linewidth}{!}{
\begin{tabular}{c|c|c}
        \toprule
        \multirow{3}{*}{Structure} &Attacked Model & Protected Model \\
        ~ & (w/ adv. training) & (w/o adv. training) \\
        ~ & Succ. Rate (\%) & Succ. Rate (\%) \\ 
        \midrule
        R50 & 99.98 $\rightarrow$ \textbf{97.09} & 26.87 $\rightarrow$ \textbf{0.02} \\
        \midrule
        V16 & 97.30 $\rightarrow$ \textbf{73.04} & 31.07 $\rightarrow$ \textbf{0.16} \\
        \midrule 
        D121 & 95.89 $\rightarrow$ \textbf{96.33} & 24.96 $\rightarrow$ \textbf{0.01}\\
        \bottomrule
  \end{tabular}
  }
  } \caption{Attack success rates of models w/ and w/o adversarial protection under the white-box setting}
\label{table:defense-model-attack}
\end{table}
\begin{table}[ht]
\centering

\subfloat[][Cifar10]{
\label{table:grey-box-cifar10-same-arch}
\resizebox{0.9\linewidth}{!}{
\begin{threeparttable}
\begin{tabular}{c|c|c|c}
  \toprule
\multicolumn{2}{c|}{Attacked Model} & \multicolumn{2}{c}{Protected Model} \\
Structure & Succ. Rate (\%) & Structure & Succ. Rate (\%) \\ 
    \midrule
 R50 & 99.61 $\rightarrow$ \textbf{99.44} & V16 & 74.78 $\rightarrow$ \textbf{5.75} \\
 R50 & 99.61 $\rightarrow$ \textbf{99.28} & D121 & 81.04 $\rightarrow$ \textbf{11.15}\\
 V16 & 97.98 $\rightarrow$ \textbf{91.67} & R50 & 43.18 $\rightarrow$ \textbf{2.99}\\
 V16 &97.98 $\rightarrow$ \textbf{91.71} & D121 &47.65 $\rightarrow$ \textbf{10.29}\\
 D121 & 99.15 $\rightarrow$ \textbf{94.16} & R50 & 80.70 $\rightarrow$ \textbf{5.26}\\
 D121 & 99.15 $\rightarrow$ \textbf{96.97} & V16 & 82.96 $\rightarrow$ \textbf{8.33}\\
 \bottomrule
  \end{tabular}
  \end{threeparttable}
  }
  } \\
\subfloat[][ImageNet]{
\label{table:grey-box-ImageNet-same-arch}
\resizebox{0.9\linewidth}{!}{
\begin{threeparttable}
\begin{tabular}{c|c|c|c}
  \toprule
\multicolumn{2}{c|}{Attacked Model} & \multicolumn{2}{c}{Protected Model} \\
Structure & Accuracy & Structure & Accuracy \\ 
    \midrule
 R50 & 99.87 $\rightarrow$ \textbf{99.28} & V16 & 50.28 $\rightarrow$ \textbf{26.34} \\
 R50 & 99.87 $\rightarrow$ \textbf{91.62} & D121 & 81.50 $\rightarrow$ \textbf{31.17}\\
 V16 & 99.65 $\rightarrow$ \textbf{97.67} & R50 & 37.93 $\rightarrow$ \textbf{22.48}\\
 V16 & 99.65 $\rightarrow$ \textbf{88.07} & D121 & 60.40 $\rightarrow$ \textbf{28.58}\\
 D121 &99.78 $\rightarrow$ \textbf{98.49} & R50 & 84.05 $\rightarrow$ \textbf{20.61}\\
 D121 & 99.78 $\rightarrow$ \textbf{99.71} & V16 &86.55 $\rightarrow$ \textbf{27.05}\\
 \bottomrule
  \end{tabular}
  \end{threeparttable}
  }
  }
  \caption{Gray-box attacks: shadow model has the same structure as the protected model}
\label{table:grey-box-attack with same arch}
\end{table}
\begin{table}[ht]
\centering

\subfloat[][Cifar10]{
\label{table:grey-box-cifar10}
\resizebox{0.99\linewidth}{!}{
\begin{tabular}{c|c|c|c|c}
  \toprule
\multicolumn{2}{c|}{Attacked Model} & \multicolumn{3}{c}{Protected Model} \\
\multirow{2}{*}{Structure} & \multirow{2}{*}{Succ. Rate (\%)} & Original & Shadow & \multirow{2}{*}{Succ. Rate (\%)} \\
& & Structure & Structure & \\
    \midrule
 R50 & 99.61 $\rightarrow$ \textbf{99.48} & V16 & V11 & 74.78 $\rightarrow$ \textbf{41.82} \\
 R50 & 99.61 $\rightarrow$ \textbf{99.53} & V16 & V19 & 74.78 $\rightarrow$ \textbf{30.74} \\
 R50 & 99.61 $\rightarrow$ \textbf{99.36} & V16 & V11\&V19 & 74.78 $\rightarrow$ \textbf{20.03}\\
 V16 & 97.98 $\rightarrow$ \textbf{96.77} & R50 & R18 & 43.18$\rightarrow$ \textbf{19.65}\\
 V16 & 97.98 $\rightarrow$ \textbf{96.87} & R50 & R34 & 43.18$\rightarrow$ \textbf{22.49}\\
 V16 & 97.98 $\rightarrow$ \textbf{95.24} & R50 & R18\&R34 & 43.18$\rightarrow$ \textbf{12.65}\\
 \bottomrule
  \end{tabular}
  }
  } \\
\subfloat[][ImageNet]{
\label{table:grey-box-ImageNet}
\resizebox{0.99\linewidth}{!}{
\begin{tabular}{c|c|c|c|c}
  \toprule
\multicolumn{2}{c|}{Attacked Model} & \multicolumn{3}{c}{Protected Model} \\
\multirow{2}{*}{Structure} & \multirow{2}{*}{Succ. Rate (\%)} & Original & Shadow & \multirow{2}{*}{Succ. Rate (\%)} \\
& & Structure & Structure & \\
    \midrule
 R50 & 99.87$\rightarrow$ \textbf{95.50} & V16 & V11 & 50.28 $\rightarrow$ \textbf{39.92} \\
 R50 & 99.87$\rightarrow$ \textbf{99.64} & V16 & V19 & 50.28 $\rightarrow$ \textbf{37.73} \\
 R50 & 99.87$\rightarrow$ \textbf{99.43} & V16 & V11\&V19 & 50.28 $\rightarrow$ \textbf{32.86} \\
 V16 & 99.65 $\rightarrow$ \textbf{99.28} & R50 & R18  & 37.93 $\rightarrow$ \textbf{32.23}\\
 V16 & 99.65 $\rightarrow$ \textbf{99.22} & R50 & R34 & 37.93 $\rightarrow$ \textbf{31.98}\\
 V16 & 99.65 $\rightarrow$ \textbf{98.92} & R50 & R18\&R34 & 37.93 $\rightarrow$ \textbf{28.51}\\
 \bottomrule
  \end{tabular}
  }
  }
  \caption{Gray-box attacks:shadow model and protected model are from the same family}
\label{table:grey-box-attack}
\end{table}
\begin{table}[ht]
\centering
\small
\subfloat[][Cifar10]{
\label{table:black-box-cifar10}
\begin{tabular}{c|c|c}
  \toprule
category& Attacked Model & Protected Model \\
  \midrule
 0 & 82.86 $\rightarrow$ \textbf{54.08} & 83.24 $\rightarrow$ \textbf{13.86}\\
 1 & 87.72 $\rightarrow$ \textbf{52.90} & 87.54 $\rightarrow$ \textbf{11.99}\\
 2 & 76.98 $\rightarrow$ \textbf{47.21} & 75.70 $\rightarrow$ \textbf{15.04}\\
 3 & 71.90 $\rightarrow$ \textbf{53.57} & 69.04 $\rightarrow$ \textbf{14.93}\\
 4 & 75.60 $\rightarrow$ \textbf{52.14} & 70.72 $\rightarrow$ \textbf{14.94}\\
 5 & 75.10 $\rightarrow$ \textbf{60.27} & 72.12 $\rightarrow$ \textbf{14.46}\\
 6 & 83.58 $\rightarrow$ \textbf{53.28} & 79.80 $\rightarrow$ \textbf{13.77}\\
 7 & 85.38 $\rightarrow$ \textbf{53.90} & 82.56 $\rightarrow$ \textbf{14.05}\\
 8 & 89.12 $\rightarrow$ \textbf{53.59} & 86.90 $\rightarrow$ \textbf{15.05}\\
 9 & 80.18 $\rightarrow$ \textbf{48.48} & 85.52 $\rightarrow$ \textbf{18.19}\\
 \bottomrule
  \end{tabular}
  } \\
\subfloat[][ImageNet]{
\label{table:black-box-ImageNet}
\begin{tabular}{c|c|c}
  \toprule
category& Attacked Model & Protected Model \\
    \midrule
 0 & 53.22 $\rightarrow$ \textbf{47.24} & 61.39 $\rightarrow$ \textbf{23.31}\\
 1 & 49.73 $\rightarrow$ \textbf{51.99} & 48.08 $\rightarrow$ \textbf{18.20}\\
 2 & 65.52 $\rightarrow$ \textbf{54.29} & 58.45 $\rightarrow$ \textbf{21.92}\\
 3 & 49.73 $\rightarrow$ \textbf{44.21} & 48.68 $\rightarrow$ \textbf{19.24}\\
 4 & 47.38 $\rightarrow$ \textbf{49.10} & 51.10 $\rightarrow$ \textbf{17.77}\\
 5 & 58.03 $\rightarrow$ \textbf{52.91} & 46.99$\rightarrow$ \textbf{18.97}\\
 6 & 62.31 $\rightarrow$ \textbf{57.31} & 58.33 $\rightarrow$ \textbf{25.05}\\
 7 & 48.46 $\rightarrow$ \textbf{42.91} & 59.32 $\rightarrow$ \textbf{18.28}\\
 8 & 55.89 $\rightarrow$ \textbf{50.45} & 48.49 $\rightarrow$ \textbf{19.49}\\
 9 & 55.11 $\rightarrow$ \textbf{57.91} & 66.66 $\rightarrow$ \textbf{21.44}\\
 \bottomrule
  \end{tabular}
  }
\caption{Results of black-box attacks using Transferability-aware Classifier}
\label{table:Black-box-attack}
\end{table}


\section{Evaluation}
\label{sec:eval}
In this section, we will first introduce the experimental configuration. Then we will evaluate the effectiveness of our methods on the whitebox, graybox and blackbox settings, respectively.

\subsection{Experimental Configuration}\label{sec:setup}

\textbf{Datasets and DL models. }
We mainly consider three CNN models with different network structures: Resnet50 \cite{He2015Resnet}, VGG16 \cite{Simonyan2014VGG} and Densenet121 \cite{Huang2017Densenet}. We select two representative datasets for image classification: Cifar10 \cite{cifar10} and ImageNet \cite{ImageNet_cvpr09}. Other deep learning models and datasets can be attacked in the same way. Cifar10 dataset contains 50,000 training samples and 10,000 testing samples of 10 classes. Each image has the size of $32 \times 32 \times 3$. For ImageNet, we select ILSVRC-2012 dataset, which contains more than 1.2 million training samples, 50,000 validation samples and 10,000 testing samples of 1,000 classes. The original images vary in size, and we reshape them to $224 \times 224 \times 3$. 

\textbf{Adversarial examples generation.}
We adopt Basic Iterative Method (BIM) \cite{kurakin2016adversarial} to generate adversarial examples. It is the basis of new transferable adversarial attack techniques, \eg, MI-FGSM \cite{Dong2018Boosting}, TI-FGSM \cite{dong2019evading} DI$^2$-FGSM \cite{xie2019improving}. We set a high noise level ($\epsilon=32$) to increase the transferability of the generated adversarial examples. This can better reflect the effectiveness of our solutions. 

We implemented our attacks with PyTorch 1.3.1. All the experiments in this paper were conducted on a server equipped with 2 Intel Xeon Sliver 4214 CPUs and 1 NVIDIA 2080Ti GPU. 


\subsection{White-box Attacks}

\textbf{Different structures.}
As the first study, we consider the attacked and protected models with different network structures. Table~\ref{table:white-box-attack} shows the attack success rates of the attacked and protected models with combinations of various structures. We abbreviate Resnet50 as R50, VGG16 as V16, and Densenet121 as D121. For each entry, the left number denotes the success rates of the two models when we generate AEs from the attacked model and transfer them to the protected model (baseline). The right number (bold) denotes the success rates after we apply RLFE. 

We get three observations from this table. First, RLFE gives very satisfactory results: it can significantly decrease the attack success rates of protected models, while still maintaining high attack impacts on attacked models. Second, the attack results are better for ImageNet than Cifar10. The non-transferability is easier to achieve for more complicated images, as the search space is larger. Third, RLFE can effectively handle multiple models. We included two different models in the protected model set, and the adversary can generate the expected adversarial examples for both of the two models with very low success rates.

\textbf{Adversarial training.}
One possible defense against adversarial attacks is adversarial training \cite{kurakin2016adversarial, huang2015learning, shaham2018understanding}, where AEs are used with normal samples together to train DNN models to recognize and correct malicious samples. However, the new model is still vulnerable to white-box attacks as the adversary can generate new AEs based on this model. We further show that with our RLFE method, the adversary can generate AEs to only attack the models with adversarial training protection, and protect the ones without AE protection.

Table~\ref{table:defense-model-attack} shows the transferability across attacked and protected models. We can see that even after the model is retrained with AEs, we can still generate new AEs to attack the new model with very high success rate. These new AEs have around $20\%$ to $30\%$ chances to affect the original model as well. After we apply RLFE, we can decrease the success rate on the original models to below 0.5\%, without affecting the attack effects on the retrained models. This confirms that RLFE has the capability to attack the models with defenses and protect the models without defenses.

\subsection{Gray-box Attacks}\label{sec:grey Box Attack}

\textbf{Same model structure.}
Next we consider the gray-box attacks. We assume the adversary has white-box access to the attacked model, but gray-box access to the protected model. He needs to train a shadow model with the same network structure. In our experiment, we reset the logits layer's weights and retrain the model for 100 epochs as the shadow model. Then the adversary can generate non-transferable AEs using RLFE. Table \ref{table:grey-box-attack with same arch} reports the attack success rate of the attacked model, and the protected model (not the shadow model). 

We can observe that this RLFE with shadow model construction can significantly reduce the non-transferability as expected. The attack effects are not as good as the white-box attacks, as we generate the AEs from the shadow model instead of the protected model. Besides, we also find that the attack effects on Cifar10 is better than that on ImageNet. The reason is that the ImageNet models are more complicated than the Cifar10 models, so the difference between constructed shadow and original models is larger, making it harder to precisely protect the original models against AEs.

\textbf{Same model family.}
We consider another case of gray-box scenario: the adversary does not know the network structure of the protected model, but he knows which family this model originates from. Then he can pick a random structure from the same family to train the shadow model. In our experiment, when the protected model is VGG16BN, the adversary can use its variant, VGG11BN or VGG19BN for shadow model construction. When the protected model is Resnet50, the adversary can choose Resnet18 or Resnet34 from the same family. Table \ref{table:grey-box-attack} displays the attack results. 

We can still observe the effectiveness of our solutions for both high success rate of attacked models, and low success rate of protected models. The results are worse than the attacks with the same network structure. This is expected as different structures can give larger discrepancy between the shadow and original models, even they are from the same family. 

\begin{figure}[!t]
  \centering
  \subfloat[Cifar10]{\includegraphics[width=0.9\columnwidth]{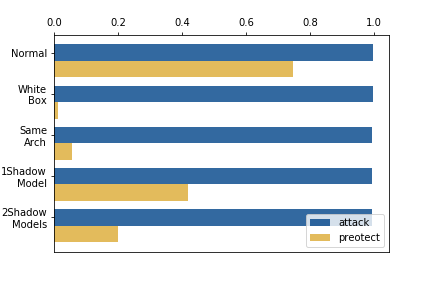}}
  \\
  \subfloat[Imagenet]{\includegraphics[width=0.9\columnwidth]{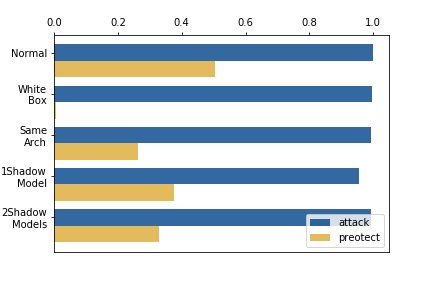}}
  \caption{The summary of Reversed Loss Function Ensemble Attacking Method. In this figure, attacked model is Resnet50, and protected model is VGG16BN. The attack success rates in different scenes have different degrees of decrease.}
  \label{img:rlfe summary}
\end{figure}

\paragraph{Qualitative Results}
We show the qualitative results in Fig.~\ref{img:rlfe summary}. \textit{Normal} represents the normal attack on Resnet50, we could find that there is a high success rate of transfer attack to VGG16. \textit{WhilteBox} shows the white-box attack on these two models, which could achieve much lower success rate on VGG16.
\textit{SameArch} means that we trained a new model but with the same architecture (i.e., VGG16), the success rate is a little higher than \textit{WhilteBox}. \textit{1ShadowModel} and  \textit{2ShadowModels} represent that we trained shadow models with one or two models with different architectures (i.e., VGG19,  VGG11\&VGG19), respectively. The overall results show that our method is effective in generating adversarial examples with nn-transferability.

\subsection{Black-box Attacks}\label{section:black box attack}

We evaluate the black-box attacks with the Transferability Classification method. We set Densenet121 as the attacked black-box model and VGG16BN as the protected black-box model. The adversary adopts Resnet50 as the assisted white-box model to generate AEs and train the TC. As mentioned in Section~\ref{section Noise-based Transfer Classifier}, the classifiers are based on the Resnet18 structure. We train one classifier for each class. 

Table~\ref{table:Black-box-attack} lists the attack results. We show all 10 classes of Cifar10 and the first 10 classes of ImageNet. We observe that the success rate of the attacked model has certain decrease compared to white-box and gray-box attacks. This is because the adversary cannot access the attacked model directly. He has to identify the perturbation from the TC. This is already a good attack performance for black-box setting. Besides, the success rate of the protected model is reduced a lot, and there are big gaps between the attacked model and protected model. This proves the effectiveness of our Transferability Classification method, and the possibility of non-transferability adversarial attacks even all the models are black-box access to the adversary.

\section{Conclusions}\label{sec:concl}

In this paper, we propose a new form of non-transferable adversarial examples, which can accurately attack a set of models without compromising another sets of models, controlled by the adversary. We introduce two methods to generate such samples under different attack scenarios. The first one is gradient-based approach: we identify a new loss function considering both attacked and protected models. Then AEs can be generated by maximizing this function and calculating its gradients. The second one is perturbation-based approach: we build a transferability-aware classifier to predict the attack results of adversarial perturbations as well as generate desired AEs. Our methods can efficiently produce accurate AEs with the specified non-transferability. They are generous and can be integrated with existing AE generation approaches. 
In future, we will use our non-transferable adversarial examples as a new kind of mutation for facilitating DNN testing \cite{issta19_deephunter,issre18_mutation,ase18_gauge,saner19_deepct}. Moreover, it will be worthwhile to further explore how the non-transferable adversarial examples can be obtained for other non-traditional adversarial attack modalities such as \cite{arxiv20_cosal,arxiv20_retinopathy,arxiv20_xray,arxiv20_advrain,guo2020spark,acmmm20_amora,guo2020abba,cheng2020pasadena}.

\footnotesize
\bibliography{ref}

\end{document}